\documentclass[preprint,12pt]{elsarticle}

\usepackage{amssymb}
\usepackage{amsmath}
\usepackage{url}
\usepackage{booktabs}
\usepackage{array}
\usepackage[version=4]{mhchem}
\usepackage{algorithm}
\usepackage{algpseudocode}
\usepackage{threeparttable}
\usepackage{tabularx}
\usepackage{chemformula}
\usepackage{hyperref}
\usepackage{xurl}
\usepackage{etoolbox}
\usepackage{xcolor}
\AtBeginEnvironment{algorithm}{\footnotesize}
\makeatletter
\renewcommand{\fnum@algorithm}{\footnotesize\bfseries Algorithm~\thealgorithm}
\makeatother
\DeclareMathOperator*{\argmin}{arg\,min}

\biboptions{square,comma,numbers,sort&compress}

\newcounter{bla}

\journal{Computer Physics Communications}

\begin{document}

\begin{frontmatter}

\title{\textit{VacHopPy}: A Python package for vacancy hopping analysis based on molecular dynamics simulations}

\author[label1,label2]{Taeyoung Jeong}
\author[label1,label2]{Kun Hee Ye}
\author[label1,label2]{Seungjae Yoon}
\author[label1,label2]{Dohyun Kim}
\author[label1,label2]{Yunjae Kim}
\author[label2]{Cheol Seong Hwang}
\author[label1]{Jung-Hae Choi\corref{cor1}}
\cortext[cor1]{Email: choijh@kist.re.kr}

\affiliation[label1]{organization={Electronic and Hybrid Materials Research Center, Korea Institute of Science and Technology},
             city={Seoul},
             postcode={02792}, 
             country={Korea}}
\affiliation[label2]{organization={Department of Materials Science and Engineering and Inter-University Semiconductor Research Center, Seoul National University},
             city={Seoul},
             postcode={08826},
             country={Korea}}

\begin{abstract}
Multiscale modeling, which integrates material properties from \textit{ab initio} calculations into continuum-scale simulations, is a promising strategy for optimizing semiconductor devices. However, a key challenge remains: while \textit{ab initio} methods provide diffusion parameters specific to individual migration paths, continuum equations require a single effective set of parameters that captures the overall diffusion behavior. To address this issue, we present \textit{VacHopPy}, an open-source Python package for vacancy hopping analysis based on molecular dynamics (MD). \textit{VacHopPy} extracts an effective set of hopping parameters, including hopping distance, hopping barrier, number of effective paths, correlation factor, and attempt frequency, by statistically integrating energetic, kinetic, and geometric contributions across all paths. It also includes tools for tracking vacancy trajectories and for detecting phase transitions during MD simulations. The applicability of \textit{VacHopPy} is demonstrated in three representative materials: face-centered cubic Al, rutile \ch{TiO2}, and monoclinic \ch{HfO2}. The extracted effective parameters reproduce temperature-dependent diffusion behavior and are in good agreement with previous experimental data. Provided in a simplified form, these parameters are well suited for continuum-scale models and remain valid over a wide temperature range spanning several hundred kelvins. Furthermore, \textit{VacHopPy} inherently accounts for anisotropy in thermal vibrations, a factor often overlooked, making it suitable for simulating diffusion in complex crystals. Overall, \textit{VacHopPy} establishes a robust bridge between atomic- and continuum-scale models, enabling more reliable multiscale simulations.
\\

\noindent \textbf{PROGRAM SUMMARY}

\begin{small}
\noindent
{\em Program Title:} \textit{VacHopPy}  \\
{\em Developer's repository link:} \url{https://github.com/TY-Jeong/VacHopPy} \\
{\em Licensing provisions:} MIT License  \\
{\em Programming language:} Python                                  \\
{\em Supplementary material:} Supplementary Figures (S1–S11), Supplementary Tables (S1–S6), and Supplementary Notes (1–4) are provided in a separate PDF file. \\
{\em Nature of problem:} For modeling of vacancy-mediated diffusion, \textit{ab initio} calculations provide path-specific diffusion parameters that are not directly compatible with continuum-scale models, which typically require a single set of effective parameters. Such incompatibility poses a significant challenge in accurately integrating atomistic diffusion behavior into multiscale simulation frameworks, particularly when multiple hopping paths exist in a material system.\\
{\em Solution method:} Vacancy trajectories are identified from MD simulations by analyzing time-averaged atomic forces and positions, enabling robust tracking of vacancy hopping events despite thermal fluctuations. From these trajectories, path-dependent energetic, kinetic, and geometric contributions are statistically integrated to construct a single set of effective hopping parameters. These effective parameters are formulated in a simplified and material-independent form, making them directly compatible with continuum-scale models without further modification.\\
{\em Additional comments:} At the time of writing, the latest version of \textit{VacHopPy} is 3.1.0. Due to the large size of the example files, a separate download link for them is provided in the \textit{VacHopPy} Documentation (\url{https://vachoppy.readthedocs.io}).\\
\\

\end{small}
   \end{abstract}
\end{frontmatter}

\newpage
\section{Introduction}
\label{sec1}

Multiscale modeling has emerged as a promising solution for optimizing semiconductor devices, including dopant engineering, interface control, and electrochemical reaction modeling~\cite{ref1,ref2,ref3,ref4,ref5}. A typical approach involves parameterizing continuum-scale models (e.g., device-level simulations) using material properties derived from atomic-scale models (e.g., \textit{ab initio} calculations)~\cite{ref6,ref7,ref8}. In this framework, material properties obtained from atomic-scale models serve as input parameters for continuum-scale models, effectively bridging microscopic physics (e.g., atom–atom interactions) into macroscopic behavior. This integration enhances the predictive capabilities of the simulations and allows for the exploration of device performance under extreme or experimentally inaccessible conditions~\cite{ref9,ref10,ref11,ref12}.

Vacancy hopping is a predominant atomic transport mechanism in crystalline systems, as vacancies provide low-energy pathways for atomic migration. This mechanism is frequently assumed in device-level models where vacancy or ion transport plays a crucial role, such as resistive random-access memory (RRAM) and solid-state electrolytes~\cite{ref19,ref20,ref21,ref22}. The \textbf{key hopping parameters} governing atomic transport through this mechanism are summarized in Table~\ref{tab:hop_params}. These parameters can be categorized into two groups: \textbf{path parameters}, which characterize individual hopping paths, and \textbf{transport parameters}, which describe the overall transport behavior. Path parameters are typically calculated using the nudged elastic band (NEB) method, a well-established \textit{ab initio} approach for analyzing atomic migration~\cite{ref13}. On the other hand, transport parameters are commonly determined using alternative approaches, such as molecular dynamics (MD) or kinetic Monte Carlo (KMC) simulations, to capture long-term transport dynamics~\cite{ref14,ref15,ref16,ref17,ref18}.

Several software packages—such as MDAnalysis, Kinisi, GEMDAT, and Pymatgen—have been developed for analyzing MD simulations~\cite{mdanalysis,kinisi,gemdat,pymatgen}.  MDAnalysis provides a comprehensive toolkit for structural and dynamical trajectory analysis, while Kinisi focuses on accurate diffusivity ($D$) estimation using probabilistic Bayesian regression. However, these packages are primarily designed for general-purpose analysis and do not extract mechanism-specific parameters beyond $D$. Consequently, a dedicated code specialized for vacancy-mediated hopping is required to determine a complete set of hopping parameters.

Estimating hopping parameters for multiscale modeling is nontrivial. Path parameters are inherently path-dependent, meaning that systems with multiple hopping paths yield diverse sets of path parameters. In contrast, continuum-scale models commonly employ simplified equations that overlook this path diversity, thereby requiring a single representative set of path parameters~\cite{ref20,ref23,ref24,ref25}. Accordingly, the coexistence of multiple hopping paths introduces ambiguity in selecting appropriate input parameters for continuum-scale models, potentially leading to inaccuracy.

\begin{table}[htbp]
  \renewcommand{\arraystretch}{1.2}
  \setlength{\tabcolsep}{4pt}
  \caption{Key hopping parameters governing atomic transport via the vacancy hopping mechanism. Path parameters characterize individual hopping paths, while transport parameters describe the overall atomic transport behavior.}
  \label{tab:hop_params}
  \centering
  \footnotesize
  \begin{tabular}{@{}>{\centering\arraybackslash}m{1.2cm} p{5.6cm}@{}}
    \toprule
    \multicolumn{2}{c}{\rule{0pt}{1.1em}\textbf{Hopping parameter}} \\
    \midrule
    \multicolumn{2}{c}{\textit{Path parameters}} \\
    \cmidrule(l){1-2}
    $a$      & Hopping distance (Å) \\
    $E_a$    & Hopping barrier (eV) \\
    $z$      & Number of equivalent paths \\
    $\nu$    & Jump attempt frequency (THz) \\
    \addlinespace
    \cmidrule(l){1-2}
    \multicolumn{2}{c}{\textit{Transport parameters}} \\ 
    \cmidrule(l){1-2}
    $D$      & Diffusivity ($\mathrm{m^2/s}$) \\
    $\tau$   & Vacancy residence time (ps) \\
    $f$      & Correlation factor \\
    \bottomrule
  \end{tabular}
\end{table}

One potential solution is to extend continuum-scale equations to account for multiple hopping paths; however, this approach can significantly increase model complexity and may compromise computational efficiency. Moreover, any change in material composition or structure necessitates re-parameterization of the equations to reflect the corresponding hopping characteristics. Alternatively, a more practical approach is introducing \textbf{effective hopping parameters}, which are expressed in a consistent form, independent of material-specific characteristics. In this strategy, multiple sets of path parameters are consolidated into a single set of effective path parameters, while the corresponding transport parameters are approximated by effective transport parameters represented in a simplified form (e.g., Arrhenius expression). Consequently, this strategy enables the seamless integration of material properties derived from atomic-scale models into continuum-scale models without modifications while reducing the complexity of the parameter space. However, a systematic method for determining these effective hopping parameters has not yet been established and remains a critical challenge in multiscale modeling.

To address this challenge, we present \textbf{\textit{VacHopPy}}, an open-source Python package for analyzing vacancy transport from MD simulations in systems dominated by the vacancy hopping mechanism. Its key features include (1) the calculation of effective hopping parameters, (2) the tracking of vacancy trajectories in MD data, and (3) the assessment of lattice stability and detection of phase transitions. The third feature identifies unstable systems in which lattice sites fail to be maintained during MD simulations, rendering vacancy definitions ambiguous and results unreliable. A comparison of the effective hopping parameters extractable with \textit{VacHopPy} versus other packages is summarized in Table S1 of the Supplementary Material. These effective parameters incorporate thermodynamic, kinetic, and geometric factors of each hopping path and are expressed in a simplified form suitable for direct use in continuum-scale models. To demonstrate its applicability, this package is validated using three representative systems: face-centered cubic (fcc, Fm$\bar{3}$m) Al, rutile (P4$_2$/mnm) \ce{TiO2}, and monoclinic (\textit{mono}, P2$_1$/c) \ce{HfO2}, materials of interest for next-generation memory devices. By providing a robust and systematic framework for extracting effective hopping parameters, \textit{VacHopPy} effectively facilitates the seamless integration of atomic-scale insights into continuum-scale simulations, playing a crucial role in multiscale modeling.

\section{\textit{VacHopPy} background}
\label{sec2}
Here, the theoretical background for the core functionalities of \textit{VacHopPy} is introduced. First, the algorithm for determining vacancy trajectories in MD simulations is described, followed by the formulation of effective hopping parameters based on the quantities derived from these trajectories. All notations used in the main text are summarized in Table~S2 of the Supplementary Material.

\subsection{Vacancy trajectory determination}
\label{sec2.1}
In MD simulations, only atomic positions are explicitly available; thus, vacancy locations must be inferred. A simple approach is to assign each atom to its nearest lattice site and designate any unoccupied site as a vacancy. However, accurate site assignment should not rely solely on proximity. Instead, the position of the atom relative to the transition state (TS) should also be considered.

\begin{figure}[t]
\centering
\includegraphics{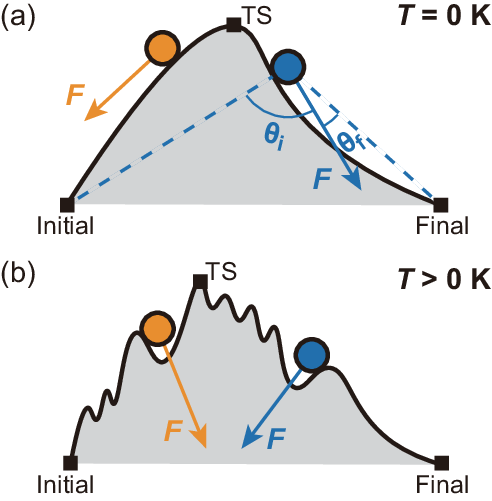}
\caption{(a) PES diagram at $T=0~K$. $\mathbf{F}$ acting on the atom before passing the TS (orange) is directed toward the initial site ($\cos \theta_i > \cos \theta_f$), where $\theta_i$ and $\theta_f$ denote the angles between the $\mathbf{F}$ and the directions toward the initial and final sites, respectively. In contrast, $\mathbf{F}$ acting on the atom after crossing the TS (blue) is directed toward the final site ($\cos \theta_i < \cos \theta_f$). (b) PES diagram at $T>0~K$. The PES is perturbed by random thermal fluctuations, making it challenging to determine the relative position of an atom to the TS based on instantaneous $\mathbf{F}$.}
\label{fig1}
\end{figure}

At 0~K, this relative position can be estimated from the force vector ($\mathbf{F}$) acting on the atom. As shown in the potential energy surface (PES) in Fig.~\ref{fig1}(a), an atom that has not passed the TS (orange) experiences a $\mathbf{F}$ directed toward the initial site, satisfying $\cos \theta_i > \cos \theta_f$, where $\theta_i$ and $\theta_f$ denote the angles between $\mathbf{F}$ and the directions to the initial and final sites, respectively. Conversely, after crossing the TS (blue), the $\mathbf{F}$ points toward the final site, yielding $\cos \theta_i < \cos \theta_f$.

At finite temperature $T>0~K$, however, thermal fluctuations perturb the PES, potentially misaligning the direction of $\mathbf{F}$. As shown in Fig.~\ref{fig1}(b), such fluctuations may cause an atom before crossing the TS (orange) to experience a $\mathbf{F}$ toward the final site or an atom after crossing the TS (blue) to experience a $\mathbf{F}$ toward the initial site. These misalignments lead to incorrect site assignments, thereby reducing the reliability of vacancy trajectories. Due to this issue, the existing codes determine the site occupation based only on positional proximity rather than the TS-criterion~\cite{gemdat,pymatgen}.

To suppress the effects of thermal fluctuation, \textit{VacHopPy} implements a trajectory refinement algorithm based on a time-averaged PES, as summarized in Algorithm 1. This algorithm is based on the random nature of thermal fluctuations, which allows them to be effectively smoothed out by time averaging. To construct the time-averaged PES, the atomic coordinates ($\rho$) and $\mathbf{F}$ from MD data are averaged over a predefined time interval ($t_\text{interval}$), with each averaged segment constituting a single step. Based on these time-averaged atomic coordinates ($\rho_\text{avg}$), atoms in each step are initially assigned to their nearest lattice sites to generate a preliminary occupation map. This map is then refined using the TS criterion applied to the averaged forces ($\mathbf{F}_\text{avg}$): only hops satisfying the condition $\cos \theta_i < \cos \theta_f$ are considered valid, while the others are discarded. The occupation map is updated iteratively until no more invalid hops remain.

\begin{center}
\begin{minipage}{10cm}
\begin{algorithm}[H]
\footnotesize
\caption{Vacancy trajectory determination}
\textbf{Input:} MD data, $t_{\text{interval}}$ (\textit{Optional})\\
\textbf{Output:} Trajectories of atoms and vacancies
\begin{algorithmic}[1]
    \State \textcolor{gray}{// Step 1: compute time-averaged PES}
    \State $(\rho_{\text{avg}}, \mathbf{F}_{\text{avg}}) \gets \texttt{TimeAveragedPES}(\text{MD data}, t_{\text{interval}})$
    
    \State \textcolor{gray}{// Step 2: initial occupation based on proximity}
    \State $occupation \gets \texttt{ProximityBasedOccupation}(\rho_{\text{avg}})$
    
    \State \textcolor{gray}{// Step 3: refinement using TS criterion}
    \Repeat
        \State $hopping\_sequence \gets \texttt{GetHopSequence}(occupation)$
        \State $valid\_hops \gets [\ ]$
        \State $invalid\_hops \gets [\ ]$
        \ForAll{$hop \in hopping\_sequence$}
            \State \textcolor{gray}{// $\theta_i$ and $\theta_f$ calculated from $\rho_{\text{avg}}$ and $\mathbf{F}_{\text{avg}}$}
            \If{$\cos \theta_i < \cos \theta_f$}
                \State $valid\_hops.\texttt{append}(hop)$
            \Else
                \State $invalid\_hops.\texttt{append}(hop)$
            \EndIf
        \EndFor
        \State $occupation \gets \texttt{UpdateOccupation}(valid\_hops)$
    \Until{$\texttt{length}(invalid\_hops) = 0$}
\end{algorithmic}
\end{algorithm}
\end{minipage}
\end{center}

Our TS-based refinement yields physically more reliable vacancy trajectories than proximity-only methods when an appropriate $t_\text{interval}$ is used. By default, $t_\text{interval}$ is set to the inverse of mean atomic vibrational frequency $\omega$ (\textit{i.e.}, $t_\text{interval}=1/\omega$), which effectively averages out thermal fluctuations. If not provided by the user, \textit{VacHopPy} automatically estimates $\omega$ via a Fourier transform of atomic displacements within the vibrational regime~\cite{vibration}. Practical guidance—including comparison with proximity-based method, an issue for collective atomic motions, and handling non-vacancy-mechanism migrations—is provided in the Supplementary Note~1. We note that the TS criterion assumes a sufficiently smooth PES; in systems with highly corrugated landscapes, its reliability may degrade.

Upon completion of this algorithm, \textit{VacHopPy} provides a comprehensive summary of the vacancy hopping history, including all accessible hopping paths within a given system, the full hopping sequence, and key statistical information such as the number of hopping events for each path and the residence time of vacancies at each lattice site.

\subsection{Effective hopping parameters}
\label{sec2.2}

An arbitrary lattice with $n$ symmetrically non-equivalent vacancy sites is considered, where each $i^\mathrm{th}$ site contains $m_i$ symmetrically non-equivalent hopping paths. For clarity, the $j^\mathrm{th}$ path originating from the $i^\mathrm{th}$ site is denoted as the $(i,j)$ path, and any parameter $p$ associated with this path is represented as $p_{(i,j)}$. To simplify the description of overall vacancy diffusion, a virtual lattice, termed the effective lattice, is introduced. This effective lattice is characterized by a set of effective hopping parameters, which are designed to reproduce the overall hopping behavior of the original lattice. These effective parameters are denoted with an overbar (e.g., $\bar{p}$).

\begin{table}[htbp]
  \renewcommand{\arraystretch}{1.4}
  \setlength{\tabcolsep}{4pt}
  \caption{Comparison between the original lattice and the effective lattice. $T$-dependent parameters are indicated by an asterisk.}
  \label{tab:hopping}
  \begin{center}
  \begin{threeparttable}
    \footnotesize
    \begin{tabular}{@{}>{\centering\arraybackslash}p{4.3cm} >{\centering\arraybackslash}p{4.3cm}@{}}
      \toprule
      \textbf{Original lattice} & \textbf{Effective lattice} \\
      \midrule
      \multicolumn{2}{c}{\textit{Path parameters}} \\
      \cmidrule(l{5pt}r{5pt}){1-2}
      $\left\{ a_{(i,j)} \,\middle|\, 1 \leq i \leq n,\, 1 \leq j \leq m_i \right\}$ & $\bar{a}$\tnote{*}\\
      $\left\{ E_{a,(i,j)} \,\middle|\, 1 \leq i \leq n,\, 1 \leq j \leq m_i \right\}$ & $\bar{E}_a$ \\
      $\left\{ z_{(i,j)} \,\middle|\, 1 \leq i \leq n,\, 1 \leq j \leq m_i \right\}$ & $\bar{z}$\tnote{*} \\
      $\left\{ \nu_{(i,j)} \,\middle|\, 1 \leq i \leq n,\, 1 \leq j \leq m_i \right\}$ & $\bar{\nu}$\tnote{*} \\
      \cmidrule(l{5pt}r{5pt}){1-2}
      \multicolumn{2}{c}{\textit{Transport parameters}} \\
      \cmidrule(l{5pt}r{5pt}){1-2}
      $D$\tnote{*} (multiple-exponential) & $\bar{D}$\tnote{*} (single-exponential) \\
      $\tau$\tnote{*} (multiple-exponential) & $\bar{\tau}$\tnote{*} (single-exponential) \\
      $f$\tnote{*} & $f$\tnote{*} ($=\bar{f}$) \\
      \bottomrule
    \end{tabular}
    \begin{tablenotes}
      \footnotesize \raggedleft
      \item[*] $T$-dependent parameters
    \end{tablenotes}
  \end{threeparttable}
  \end{center}
\end{table}

Table~\ref{tab:hopping} compares the original and effective lattices. In our framework, the effective lattice consists of a single hopping path characterized by a single set of effective path parameters, while the original lattice contains $\sum_{i=1}^{n}m_i$ distinct hopping paths, each with its own path parameters. In the original lattice, the $T$-dependencies of diffusivity ($D$) and vacancy residence time ($\tau$) are nontrivial, as each path has a distinct $E_{a,(i,j)}$ and contributes differently to the overall behavior~\cite{ref26}. Accordingly, $D$ and $\tau$ are expressed as multi-exponential forms, reflecting both the path-specific $T$-dependence ($\sim \exp(-E_{a,(i,j)}/k_B T)$) and their inter-path correlations. Here, $k_B$ is the Boltzmann constant. In the effective lattice, in contrast, these parameters are approximated using single exponential terms, represented as $\bar{D}$ and $\bar{\tau}$. Meanwhile, the correlation factor ($f$) is assumed to be identical in both the original and effective lattices (\textit{i.e.}, $f = \bar{f}$) and is henceforth uniformly denoted as $f$.

In this section, the effective hopping parameters are formulated based on the MD-derived quantities listed in Table~\ref{tab:aimd_quantities}, which are extracted from vacancy trajectories through MD simulations. The random walk diffusivity, representing the diffusivity for an ideal random walk, is given by $D_{\mathrm{rand}} = \sum_{(i,j)} a_{(i,j)}^{2} c_{(i,j)} / 6 t$ where $\sum_{(i,j)}$ denotes the double summation $\sum_{i=1}^n \sum_{j=1}^{m_i}$ (see Appendix A for its derivation). The descriptions for the other quantities are self-explanatory. Full derivations for the effective hopping parameters are available in Appendices A and B, while only key ideas are outlined here for clarity.

\begin{table}[htbp]
  \renewcommand{\arraystretch}{1.4}
  \setlength{\tabcolsep}{4pt}
  \caption{MD-derived quantities that can be directly determined from the vacancy trajectories derived by \textit{VacHopPy}.}
  \label{tab:aimd_quantities}
  \centering
  \footnotesize
  \begin{tabularx}{9cm}{@{}>{\centering\arraybackslash}p{1cm} X@{}}
    \toprule
    $t$ & Total simulation time \\
    $t_i$ & Total time the vacancy stayed at the $i^{\mathrm{th}}$ site \\
    $c$ & Total number of hopping events \\
    $c_{(i,j)}$ & Number of hopping events through the $(i,j)$ path \\
    $\langle \mathbf{R}^2 \rangle$ & Mean squared displacement (MSD) of the vacancy \\
    $\tau$ & Vacancy residence time ($=t/c$) \\
    $\Gamma$ & Overall hopping rate ($=c/t$) \\
    $\Gamma_{(i,j)}$ & Hopping rate through the $(i,j)$ path ($=c_{(i,j)}/t_i$) \\
    $D_{\mathrm{rand}}$ & Random walk diffusivity ($=\frac{1}{6t} \sum_{(i,j)} a_{(i,j)}^2 c_{(i,j)}$) \\
    $P_i^{\mathrm{site}}$ & Site occupation probability at the $i^{\mathrm{th}}$ site ($=t_i/t$) \\
    \bottomrule
  \end{tabularx}
\end{table}

In the effective lattice, the effective random walk diffusivity $\bar{D}_{\mathrm{rand}}$ and the effective vacancy residence time $\bar{\tau}$ are expressed as

\begin{align}
    \bar{D}_{\mathrm{rand}} &= \bar{D}_{\mathrm{rand},0} \cdot \exp\left(-\frac{\bar{E}_a}{k_{B}T}\right), 
    & \bar{D}_{\mathrm{rand},0} &= \frac{1}{6}\bar{z}\bar{a}^2\bar{\nu} \label{eq1} \\
    \bar{\tau} &= \bar{\tau}_{0} \cdot \exp\left(\frac{\bar{E}_a}{k_{B}T}\right),
    & \bar{\tau}_{0} &= \frac{1}{\bar{z}\bar{\nu}} \label{eq2}
\end{align}

where $\bar{D}_{\mathrm{rand},0}$ and $\bar{\tau}_{0}$ are the pre-exponential factors of $\bar{D}_{\mathrm{rand}}$ and $\bar{\tau}$, respectively. The factor $1/6$ in Eq.~\eqref{eq1} accounts for three-dimensional diffusion. The optimal values of $\bar{D}_{\mathrm{rand},0}$, $\bar{E}_a$, and $\bar{\tau}_{0}$ are obtained by fitting Eqs.~\eqref{eq1} and~\eqref{eq2} to the $D_{\mathrm{rand}}$ and $\tau$ values in Table~\ref{tab:aimd_quantities}, respectively:

\begin{align}
    \bar{D}_{\mathrm{rand},0}, \bar{E}_a 
    &= \argmin_{\bar{D}_{\mathrm{rand},0}, \bar{E}_a} 
    \sum_{T} \left( \ln D_{\mathrm{rand}} - \ln \bar{D}_{\mathrm{rand}} \right)^2 
    \label{eq3} \\
    \bar{\tau}_0 
    &= \argmin_{\bar{\tau}_0} 
    \sum_{T} \left( \ln \tau - \ln \bar{\tau} \right)^2 
    \label{eq4}
\end{align}

The summations are performed over all temperatures $T$ used in the MD simulations. These Arrhenius-type approximations are valid across a broad temperature range, as will be discussed in Section 5.4. By comparing the expressions for $\bar{D}_{\mathrm{rand}}$ and $\bar{\tau}$ in Eqs.~\eqref{eq1} and~\eqref{eq2}, the effective hopping distance $\bar{a}$ is given by

\begin{align}
    \bar{a} = \sqrt{6 \bar{D}_{\mathrm{rand}} \bar{\tau}}
    \label{eq5}
\end{align}

However, $\bar{D}_{\mathrm{rand}}$ corresponds to an ideal random walk, which may differ from the actual hopping process observed during MD simulations. This deviation is quantified by the correlation factor $f$, defined as

\begin{align}
    f &= \frac{D}{D_{\mathrm{rand}}} \label{eq6} \\
      &= \frac{\langle \mathbf{R}^2 \rangle}{\sum_{(i,j)} a_{(i,j)}^2 c_{(i,j)}} \label{eq7}
\end{align}

where $\langle \mathbf{R}^2 \rangle$ is the mean squared displacement (MSD) of the vacancy during the MD simulations. The second equality holds under the assumption of vacancy hopping. Typically, $f$ ranges from 0 to 1, with $f=1$ indicating fully random diffusion and $f<1$ implying correlated diffusion events. In \textit{VacHopPy}, Eq.~\eqref{eq7} is further modified using the \textit{encounter} approach to improve computational convergence, as described in Appendix~B.

The remaining effective hopping parameters, $\bar{\nu}$ and $\bar{z}$, are determined as follows:

\begin{align}
    \bar{\nu} &= \frac{\sum_{(i,j)} \nu_{(i,j)} P_{(i,j)}}{\sum_{(i,j)} P_{(i,j)}}, & P_{(i,j)} &= z_{(i,j)} P_i^{\mathrm{site}} P_{(i,j)}^{\mathrm{esc}} \label{eq8} \\
    \bar{z} &= \frac{\sum_{(i,j)} P_{(i,j)}}{\bar{P}^{\mathrm{esc}}}, & \bar{P}^{\mathrm{esc}} &= \exp\left( -\frac{\bar{E}_a}{k_B T} \right) \label{eq9}
\end{align}

Here, $P_{(i,j)}^{\mathrm{esc}}$ is the escape probability defined as $P_{(i,j)}^{\mathrm{esc}} = \exp(-E_{a,(i,j)}/k_B T)$. The term $P_i^{\mathrm{site}}$ represents the site occupation probability of the $i^{\mathrm{th}}$ site, thermodynamically proportional to $\exp(-E_{f,i}/k_B T)$, where $E_{f,i}$ is the vacancy formation energy at the $i^{\mathrm{th}}$ site. Thus, in Eqs.~\eqref{eq8} and~\eqref{eq9}, the contribution of the $(i,j)$ path decays exponentially with increasing $E_{a,(i,j)}$ or $E_{f,i}$, consistent with physical expectations. Both $\bar{\nu}$ and $\bar{z}$ are temperature-dependent since $P_{(i,j)}$ varies with temperature. Although Eq.~\eqref{eq8} can be further simplified to Eq.~\eqref{eqA9} by replacing the $\nu_{(i,j)}$ term with MD-derived quantities, it is presented here to emphasize its physical meaning. It is important to note that $P_{(i,j)}^{\mathrm{esc}}$ is not directly accessible from MD data; instead, it requires additional NEB calculations to derive $E_{a,(i,j)}$. As a result, evaluating $\bar{\nu}$ and $\bar{z}$ necessitates input from both MD and NEB calculations, whereas the other effective hopping parameters can be obtained solely from MD data.

\begin{figure}[htbp]
\centering
\includegraphics{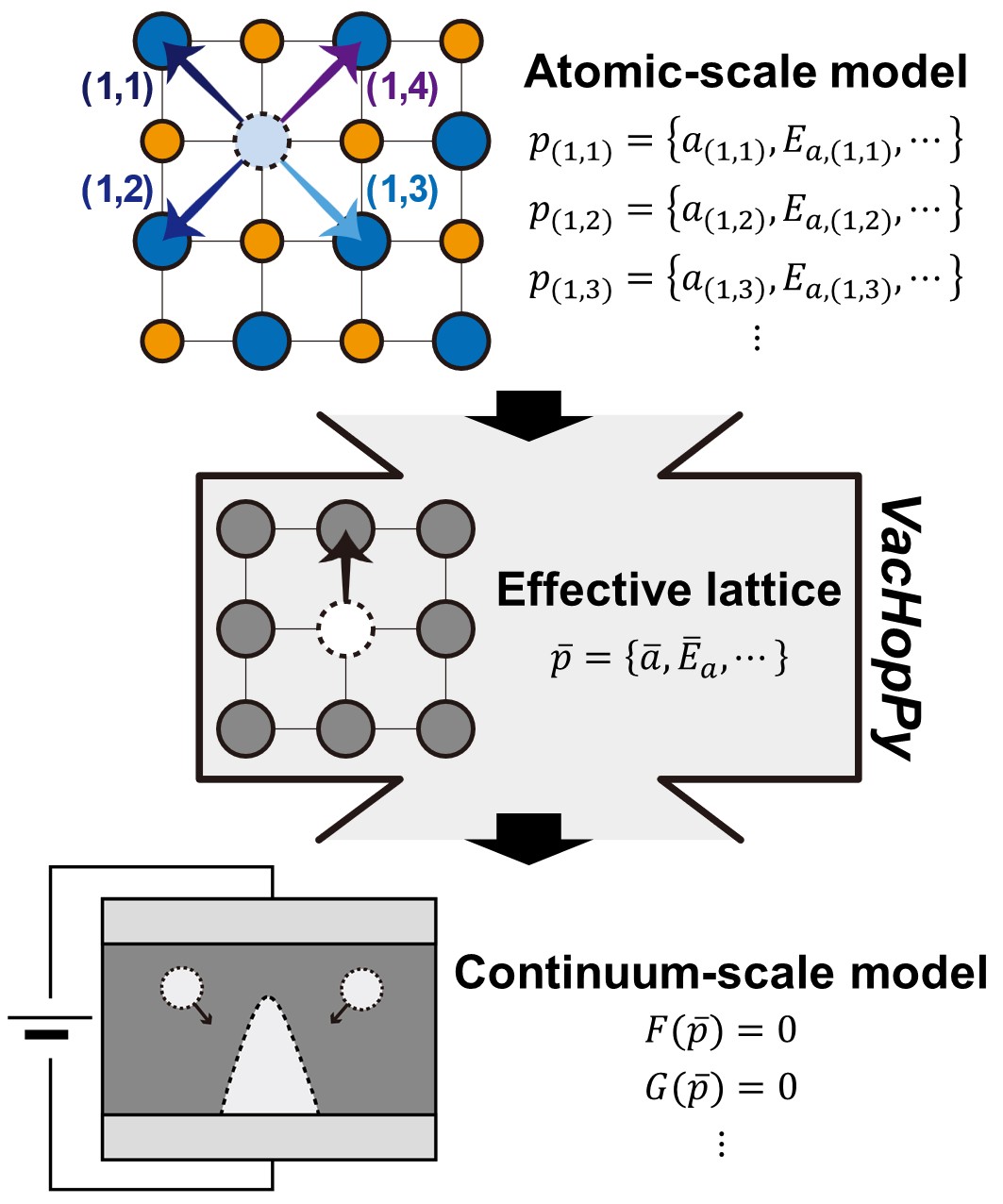}
\caption{Multiscale modeling strategy using effective hopping parameters. Solid and dashed circles represent atoms and vacancies, respectively. At the atomic scale, the hopping parameters $p_{(i,j)}$ are path-dependent, resulting in multiple distinct sets within a given system. \textit{VacHopPy} consolidates these into a single set of effective hopping parameters $\bar{p}$, expressed in a material-independent form. This abstraction resolves the ambiguity in parameter selection and enables the use of universal continuum-scale models whose governing equations are formulated in terms of $\bar{p}$ (e.g., $F(\bar{p})=0$ and $G(\bar{p})=0$).}
\label{fig2}
\end{figure}

Hereby, all effective hopping parameters listed in Table~\ref{tab:hopping} are fully defined. The multiscale modeling strategy utilizing these effective hopping parameters is schematized in Fig.~\ref{fig2}. In this framework, \textit{VacHopPy} serves as a bridge between atomic-scale models and continuum-scale models by providing a set of effective hopping parameters $\bar{p}$ as input to the latter, where the governing equations of the continuum models are formulated in terms of $\bar{p}$. Since $\bar{p}$ is formulated in a consistent and material-independent form, these models can be universally applied without modification.
 
\section{Computational details}
\label{sec3}

\textit{Ab initio} molecular dynamics (AIMD) simulations were performed to obtain atomic trajectories and the corresponding forces acting on atoms for vacancy diffusion analysis. All \textit{ab initio} calculations were conducted using the Vienna Ab initio Simulation Package (VASP) with the Perdew-Burke-Ernzerhof (PBE) exchange-correlation functional~\cite{ref27,ref28,ref29}. For fcc Al, a single neutral vacancy (\ch{V_{Al}}) was introduced, while for rutile \ce{TiO2} and \textit{mono} \ce{HfO2}, a doubly charged oxygen vacancy (\ch{V_O^{2+}}) was considered.

The AIMD simulations were conducted in canonical (NVT) ensembles using the Nosé-Hoover thermostat~\cite{ref30}. To accelerate the diffusion, simulations were conducted near the melting temperature ($T_\mathrm{melt}$) or the phase transition temperature. For Al ($T_\mathrm{melt} = 933$ K), simulations were performed at 750–900~K. For rutile \ce{TiO2} ($T_\mathrm{melt} = 2116$ K), simulations were conducted at 1700–2100 K. For \textit{mono} \ce{HfO2} ($T_\mathrm{melt} = 3031$ K), simulations were performed across two temperature ranges: low-$T$ (1600–2000 K) and high-$T$ (2200–2600 K), to account for a phase transition to the tetragonal phase (\textit{tet}, P$4_2$/nmc) occurring around 2000 K, in accordance with prior experimental and theoretical reports~\cite{ref31,voronko,luo}. Considering computational cost while mitigating interactions between periodic images, $3 \times 3 \times 3$, $2 \times 2 \times 3$, and $2 \times 2 \times 2$ supercells were employed for each system.

Path-dependent hopping barriers ($E_{a,(i,j)}$) were obtained using the climbing NEB method~\cite{ref32,ref33,ref34}. Five intermediate structures were used and relaxed until the Hellmann-Feynman forces were less than 0.05 eV/Å. For the NEB, $4 \times 4 \times 4$, $3 \times 3 \times 4$, and $3 \times 3 \times 3$ supercells were employed for Al, \ce{TiO2}, and \ce{HfO2}, respectively, which are larger than those used in the AIMD simulations. In AIMD, thermal fluctuations attenuate interactions with periodic images and allow for a broader sampling of configurations, comparable to the benefits of using a larger supercell. For comparison, $E_{a,(i,j)}$ values with respect to supercell size are shown in Table~S3 of the Supplementary Material. Additionally, all calculation conditions used in the AIMD simulations and NEB calculations are summarized in Table~S4 of the Supplementary Material.

\section{Implementation}
\label{sec4}

\begin{figure}[htbp]
  \centering
  \includegraphics[width=\textwidth]{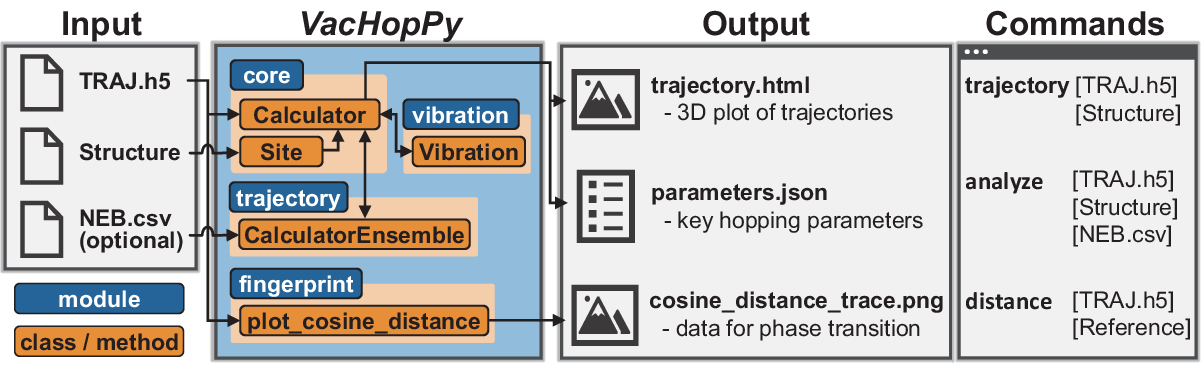}
  \caption{Representative workflow of the \textit{VacHopPy} package. For clarity, only the main API components and CLI commands are shown. Detailed descriptions of all functionalities and usage tutorials are provided in the \textit{VacHopPy} documentation (\url{https://vachoppy.readthedocs.io}).}
  \label{fig3}
\end{figure}

The latest version of \textit{VacHopPy} is available on the GitHub repository ({\small \url{https://github.com/TY-Jeong/VacHopPy}}) and can be installed via \texttt{\footnotesize pip} (e.g., \texttt{\footnotesize pip install vachoppy}). \textit{VacHopPy} supports both a command-line interface (CLI) and a Python API. The CLI allows use of the main features without writing user scripts, and the available commands are summarized in Table~S5 of the Supplementary Material. The Python API provides access to the full functionality of the package. The representative \textit{VacHopPy} workflow is summarized in Fig.~\ref{fig3}, showing only the main API components and CLI commands for clarity.

\textit{VacHopPy} reads three types of input data. The first is the \textbf{trajectory file} (e.g., \texttt{\footnotesize TRAJ.h5}), an HDF5-formatted MD trajectory generated using the \texttt{\footnotesize convert} command. This command interfaces with the Atomic Simulation Environment (ASE)~\cite{ase}, enabling the use of various ASE-compatible trajectory formats such as \texttt{\footnotesize vasprun.xml}, \texttt{\footnotesize extxyz}, and \texttt{\footnotesize LAMMPS dump} files. The second input is the \textbf{structure file}, which contains the vacancy-free crystal structure and is used to identify lattice sites and available hopping paths. The third, optional input, \texttt{\footnotesize NEB.csv}, includes \(E_{a,(i,j)}\) values for the identified hopping paths and is required to estimate \(\bar{z}\) and \(\bar{\nu}\).
 
The \texttt{\footnotesize trajectory} command generates a visualization of vacancy motion (e.g., \texttt{\footnotesize trajectory.html}), providing an intuitive view of the diffusion process. The core analysis is performed using the \texttt{\footnotesize analyze} command, which extracts the effective hopping parameters and saves them in the \texttt{\footnotesize parameters.json} file. The extractable parameters depend on the provided input data, and their correspondence is summarized in Table~S6 of the Supplementary Material. Finally, the \texttt{\footnotesize distance} command compares the MD snapshots with the reference structure and outputs their structural similarity as a function of time. The similarity is quantified by \(d_\mathrm{cos}(x)\), the cosine distance from a reference structure $x$, where smaller \(d_\mathrm{cos}(x)\) indicates higher similarity (see Appendix C for details).

The \texttt{\footnotesize trajectory} and \texttt{\footnotesize analyze} commands accept $t_\text{interval}$ as an optional argument. If not specified, $t_\text{interval}$ is automatically set to $1/\omega$, where $\omega$ is estimated by the \texttt{\footnotesize Vibration} class. For fcc Al and rutile \ce{TiO2}, the default settings ($t_\text{interval} = 0.088$ and $0.074$ ps, respectively) were used. For monoclinic \ce{HfO2}, however, a larger value of $t_\text{interval}=0.070$ ps (default 0.051 ps) was adopted to suppress large-amplitude vibrations (see Supplementary Note~1.2).

\section{Application examples and discussion}
\label{sec5}

\subsection{Fcc Al}
\label{sec5.1}

\begin{figure}[htbp]
\centering
\includegraphics{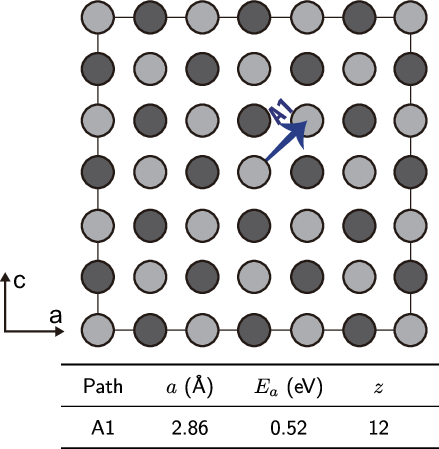}
\caption{Atomic structure and a single vacancy hopping path in fcc Al. Light and dark grey circles represent Al atoms in different layers. The characteristics of the A1 path are summarized in the accompanying table.}
\label{fig4}
\end{figure}

Fig.~\ref{fig4} shows the atomic structure of fcc Al, which features a single site and a single hopping path for \ch{V_{Al}}, named the A1 path ($n=1$ and $m_1=1$). The accompanying table summarizes the characteristics of this path. The activation energy $E_a$ of 0.52~eV, obtained using NEB calculations, agrees well with previous theoretical studies (0.52–0.63~eV) and an experimental observation (0.59~eV)~\cite{ref35,ref36}.

The effective hopping parameters for fcc Al are summarized in Table~\ref{tab:parameters_Al}. The pre-exponential factors for $\bar{D}$ ($\bar{D}_0$) and $\bar{\tau}_0$ were obtained using Eqs.~\eqref{eq1} and \eqref{eq6}, and Eq.~\eqref{eq2}, respectively. For the $T$-dependent parameters $\bar{a}$, $\bar{z}$, $\bar{\nu}$, and $f$, the mean values over the simulated temperature range are presented, with standard deviations indicated in parentheses. Their temperature dependencies are shown in Fig.~S1 of the Supplementary Material and were found to be negligible.

\begin{table}[htbp]
  \renewcommand{\arraystretch}{1.4}
  \setlength{\tabcolsep}{4pt}
  \caption{Effective hopping parameters for \ch{V_{Al}} in fcc Al ($T=750$–$900$ K), compared with their original counterparts. For $T$-dependent parameters, the mean value over the simulated temperature range is represented, with the standard deviation given in parentheses.}
  \label{tab:parameters_Al}
  \centering
  \footnotesize
  \begin{tabular}{@{}>{\centering\arraybackslash}p{1cm} >{\centering\arraybackslash}p{4cm} >{\centering\arraybackslash}p{4cm}@{}}
    \toprule
    \textbf{} & \textbf{Effective} & \textbf{Original} \\
    \midrule
    \multicolumn{3}{c}{\textit{Path parameters}} \\
    \cmidrule(lr){1-3}
    $\bar{a}$ & 2.86 ($\pm$0.0) Å & 2.86 Å* \\
    $\bar{E}_a$ & 0.52 eV & 0.52 eV* \\
    $\bar{z}$ & 12.0 ($\pm$0.0) & 12* \\
    $\bar{\nu}$ & 4.5 ($\pm$0.1) THz & $3.0\text{--}7.0$ THz~\cite{ref38} \\
    \cmidrule(lr){1-3}
    \multicolumn{3}{c}{\textit{Transport parameters}} \\
    \cmidrule(lr){1-3}
    $\bar{D}_0$ & $5.8 \times 10^{-7}$ m$^2$/s & -- \\
    $\bar{\tau}_0$ & $1.9 \times 10^{-2}$ ps & -- \\
    $f$ & 0.79 ($\pm$0.00) & 0.78~\cite{ref37} \\
    \bottomrule
  \end{tabular}
\end{table}

Since this system contains only a single hopping path, the effective lattice remains identical to the original lattice. Hence, the accuracy of the effective hopping parameters can be directly evaluated by comparing them with the original counterparts. As shown in Table~\ref{tab:parameters_Al}, $\bar{a}$, $\bar{E}_a$, and $\bar{z}$ are almost identical to those obtained from NEB calculations. Additionally, the correlation factor $f$ agrees well with the reference value reported in previous Monte Carlo simulations~\cite{ref37}. In contrast, the attempt frequency $\nu$ is known to be highly method-dependent. A previous study reported $\nu_\mathrm{A1}$ of 4.9~THz and 7.0~THz using density functional perturbation theory (DFPT) and finite difference method (FDM), respectively~\cite{ref38}, both based on transition state theory (TST), a deterministic framework using phonon properties~\cite{ref39,ref40,ref41,ref42}. Another deterministic approach, our calculation based on Eyring's formula, yields $\nu_\mathrm{A1} = 3.0$~THz (see Supplementary Note~2 of the Supplementary Material). On the other hand, \textit{VacHopPy} statistically determines $\nu_{(i,j)}$ from vacancy hopping histories, yielding $\nu_\mathrm{A1} = \bar{\nu} = 4.5$~THz, which lies within the range predicted by the TST-based calculations. This agreement further demonstrates the robustness of the \textit{VacHopPy} framework.

\subsection{Rutile \ce{TiO2}}
\label{sec5.2}

\begin{figure}[htbp]
\centering
\includegraphics{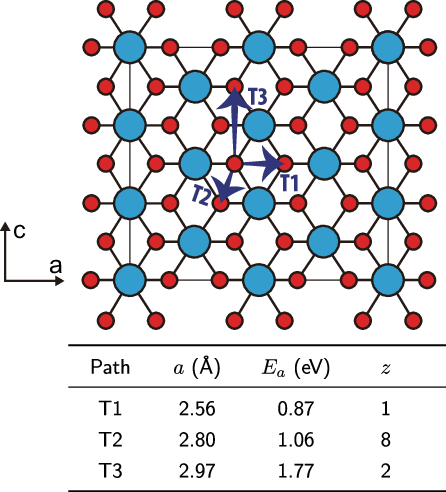}
\caption{Atomic structure and three vacancy hopping paths in rutile \ce{TiO2}. The blue and red circles represent Ti and O atoms, respectively. The characteristics of each path are summarized in the accompanying table.}
\label{fig5}
\end{figure}

Fig.~\ref{fig5} shows the atomic structure of rutile \ce{TiO2}, which features a single site and three hopping paths for \ch{V_O^{2+}} ($n=1$ and $m_1=3$). The hopping paths are named T1, T2, and T3 in ascending order of hopping distance $a$, and their characteristics are summarized in the accompanying table. Their activation energies $E_a$, 0.87, 1.06, and 1.77~eV, show good agreement with previous theoretical studies~\cite{ref43,ref44}.

The effective hopping parameters for rutile \ce{TiO2} are summarized in Table~\ref{tab:parameters_tio2}. The values of $\bar{a}$ and $\bar{E}_a$ lie between those of the T1 and T2 paths, suggesting that these two paths dominate the overall diffusion behavior. This is further supported by the vacancy hopping history: among a total of 2,455 hopping events observed across the simulated temperature range, 875, 1,578, and 2 events occurred via the T1, T2, and T3 paths, respectively. The negligible contribution of the T3 path is attributed to its significantly higher $E_{a,\mathrm{T3}}$. At the same time, the predominance of T2 arises from its coordination number $z_\mathrm{T2}$ being eight times greater than $z_\mathrm{T1}$, despite $E_{a,\mathrm{T2}}$ being higher than $E_{a,\mathrm{T1}}$.

\begin{table}[htbp]
  \renewcommand{\arraystretch}{1.4}
  \setlength{\tabcolsep}{4pt}
  \caption{Effective hopping parameters for \ch{V_O^{2+}} in rutile \ce{TiO2} ($T=1700$–$2100$ K). For $T$-dependent parameters, the mean value over the simulated temperature range is represented, with the standard deviation given in parentheses.}
  \label{tab:parameters_tio2}
  \centering
  \footnotesize
  \begin{tabular}{@{}>{\centering\arraybackslash}p{2cm} >{\centering\arraybackslash}p{4.8cm}@{}}
    \toprule
    \multicolumn{2}{c}{\textbf{Effective hopping parameter (\ce{TiO2})}}\\
    \midrule
    \multicolumn{2}{c}{\textit{Path parameters}} \\
    \cmidrule(lr){1-2}
    $\bar{a}$       & 2.71 ($\pm$0.01) Å \\
    $\bar{E}_a$     & 0.96 eV \\
    $\bar{z}$       & 6.2 ($\pm$0.1) \\
    $\bar{\nu}$     & 5.9 ($\pm$0.2) THz \\
    \cmidrule(lr){1-2}
    \multicolumn{2}{c}{\textit{Transport parameters}} \\
    \cmidrule(lr){1-2}
    $\bar{D}_0$     & $2.9 \times 10^{-7}$ m$^2$/s \\
    $\bar{\tau}_0$  & $2.7 \times 10^{-2}$ ps \\
    $f$             & 0.66 ($\pm$0.00) \\
    \bottomrule
  \end{tabular}
\end{table}

\begin{figure}[htbp]
\centering
\includegraphics[width=\textwidth]{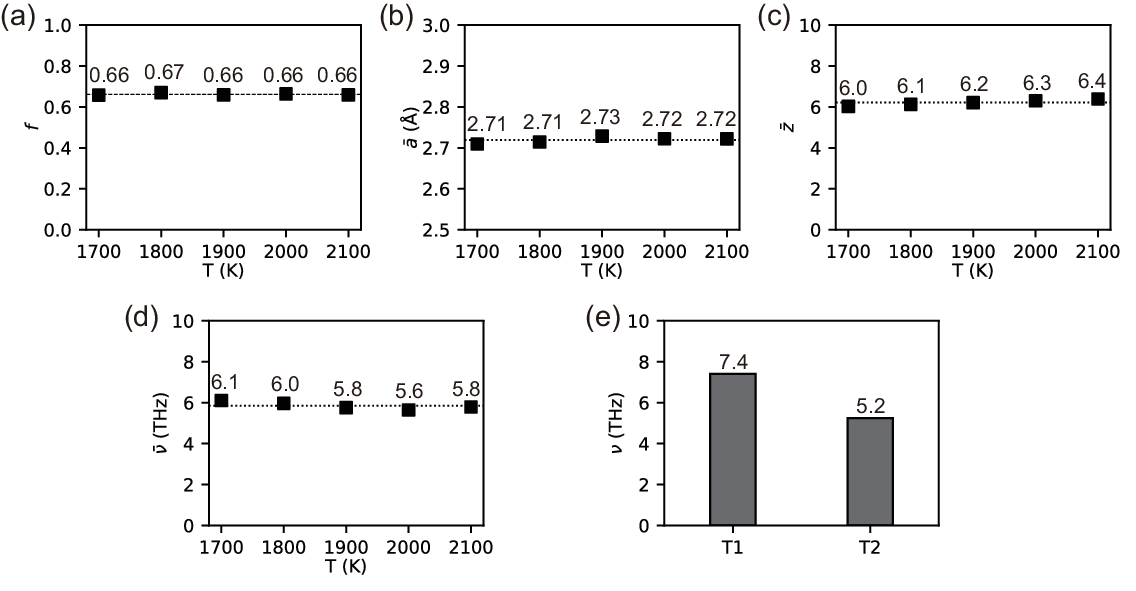}
\caption{\textit{VacHopPy} results for rutile \ce{TiO2}. (a)–(d) Temperature dependence of $f$, $\bar{a}$, $\bar{z}$, and $\bar{\nu}$, respectively. The dotted lines indicate the mean values over the simulated temperature range. (e) Averaged $\nu$ values for the T1 and T2 paths over the simulated temperature range. The T3 path is excluded due to insufficient sampling and low statistical reliability.}
\label{fig6}
\end{figure}

Figs.~\ref{fig6}(a)–(c) show the temperature dependencies of $f$, $\bar{z}$, and $\bar{\nu}$, respectively, where the mean values are indicated by dotted lines. The $f$ value remains nearly constant at 0.66 over the simulated $T$ range. In contrast, as $T$ increases, both $\bar{a}$ and $\bar{z}$ slightly increase, while $\bar{\nu}$ decreases. This trend is attributed to the increasing contribution of the T2 path at higher temperatures. At low $T$, hopping predominantly occurs via the T1 path due to its lower activation energy $E_{a,\mathrm{T1}}$. As temperature increases, thermal activation promotes hopping via the T2 path (e.g., the ratio $c_\mathrm{T2}/c_\mathrm{T1}$ increases from 1.5 at 1700 K to 1.9 at 2100 K), thereby increasing $\bar{a}$ and $\bar{z}$. Fig.~\ref{fig6}(e) shows the path-specific attempt frequencies $\nu_\mathrm{T1}$ and $\nu_\mathrm{T2}$, determined statistically as described by Eq.~\eqref{eqA17} in \textit{VacHopPy}. The value of $\nu_\mathrm{T3}$ is omitted due to insufficient sampling of hopping events via the T3 path. Notably, $\nu_\mathrm{T2}$ is smaller than $\nu_\mathrm{T1}$, which explains the observed decrease in $\bar{\nu}$ with increasing temperature as the T2 path becomes more dominant.

The higher $\nu_\mathrm{T1}$ suggests that thermal vibrations are biased toward the T1 path. Such anisotropic thermal vibration should be considered when determining the relative contribution of each path; however, this factor has often been overlooked in previous simulation studies. Instead, a typical $\nu$ value in the range of 1–10~THz was commonly assumed without considering its path dependence~\cite{ref45,ref46}. This simplification was presumably adopted because conventional TST-based methods occasionally yield $\nu$ values that are sensitive to calculation conditions. For instance, $\nu_\mathrm{T2}$ calculated using Eyring's formula was found to be approximately $10^{9}$~THz, which significantly exceeds the typical range (see Fig.~N9(d) in the Supplementary Note~2). This discrepancy is probably due to the presence of numerous imaginary phonon modes. Since Eyring's formula is highly sensitive to phonon spectra, its applicability is limited to systems that exhibit only stable phonon modes. However, in systems containing vacancies, imaginary phonons are commonly observed, thereby hindering the reliability of such calculations. In contrast, the statistical approach of \textit{VacHopPy} enables a more robust estimation of $\nu$, provided that a sufficient number of hopping events are sampled. The calculated values of $\bar{E}_a$ and $\bar{\nu}$ show good agreement with experimental data, 1.1~eV and 5~THz~\cite{ref47} and 1.15~eV and 0.3–6.3~THz~\cite{ref48}.

In summary, these results demonstrate that the effective hopping parameters from \textit{VacHopPy} properly incorporate the energetic (e.g., $E_{a,(i,j)}$), kinetic (e.g., $\nu_{(i,j)}$), and geometric (e.g., $z_{(i,j)}$) contributions of individual hopping paths. Hence, using these effective parameters as input for continuum-scale models enables accurate reflection of material-specific characteristics, thereby enhancing the predictive power and reliability of the models.

\subsection{Monoclinic \ce{HfO2}}
\label{sec5.3}

\begin{figure}[htbp]
  \centering
  \includegraphics[width=\textwidth]{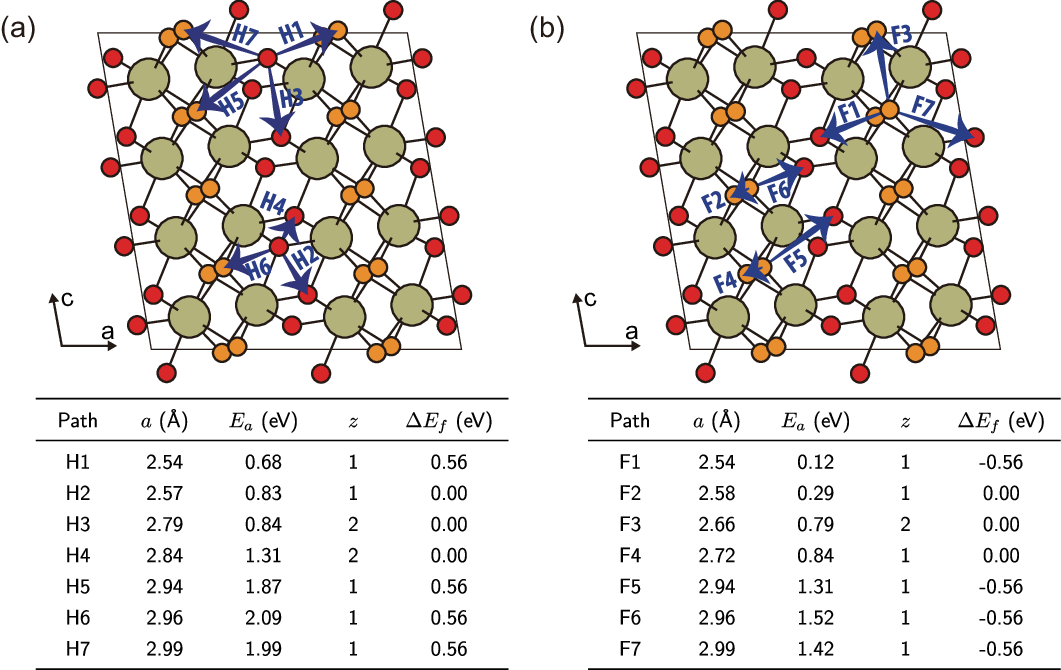}
  \caption{Atomic structure containing two symmetrically distinct oxygen vacancy sites and their associated hopping paths in \textit{mono} \ce{HfO2}. Yellow circles represent Hf atoms, while the red and orange circles indicate the 3-coordinated and 4-coordinated oxygen atoms, respectively. (a) Characteristics of the H-paths originating from the 3-coordinated site (H-site) (b) Characteristics of the F-paths originating from the 4-coordinated site (F-site). The characteristics of each hopping path are summarized in the accompanying tables. Among them, the H1-F1, H5-F5, H6-F6, and H7-F7 path pairs correspond to hopping in opposite directions, while other paths maintain their coordination numbers. In these tables, $\Delta E_f$ represents the difference in  $E_f$ between the initial and final sites of each hopping path. Under our calculation conditions, the H-site is more stable than the F-site by 0.56~eV.}
  \label{fig7}
\end{figure}

Fig.~\ref{fig7} shows the atomic structure of \textit{mono} \ce{HfO2}, which contains two symmetrically distinct sites for \ch{V_O^{2+}} ($n=2$). These sites are indicated by red circles (3-coordinated sites and denoted as H-sites) and orange circles (4-coordinated sites and denoted as F-sites). The H-site is energetically more stable for \ch{V_O^{2+}} by $\Delta E_f=0.56$~eV than the F-site. The H-site has seven paths, $m_\mathrm{H}=7$, and the corresponding H-paths are named H1, H2, \ldots, and H7 in ascending order of $a$ in Fig.~\ref{fig7}(a). The F-site also has seven paths, $m_\mathrm{F}=7$, and the corresponding F-paths are named F1, F2, \ldots, and F7 in ascending order of $a$ in Fig.~\ref{fig7}(b). The characteristics of the hopping paths are summarized in the accompanying tables of Figs.~\ref{fig7}(a) and (b), where the $E_a$ values show good agreement with previous theoretical studies~\cite{ref49}.

\subsubsection{Low-$T$ region ($T\leq2000$ K)}
\label{sec5.3.1}

From the AIMD data, a total of 4,036 hopping events were detected across the simulated $T$ range. Among them, 3,049 (76\%) events correspond to H-paths, and 987 (24\%) events correspond to F-paths. Furthermore, vacancies were found to reside approximately 24 times longer at the H-sites than at the F-sites ($t_\mathrm{H}\approx24t_\mathrm{F}$). 

\begin{figure}[htbp]
  \centering
  \includegraphics[width=\textwidth]{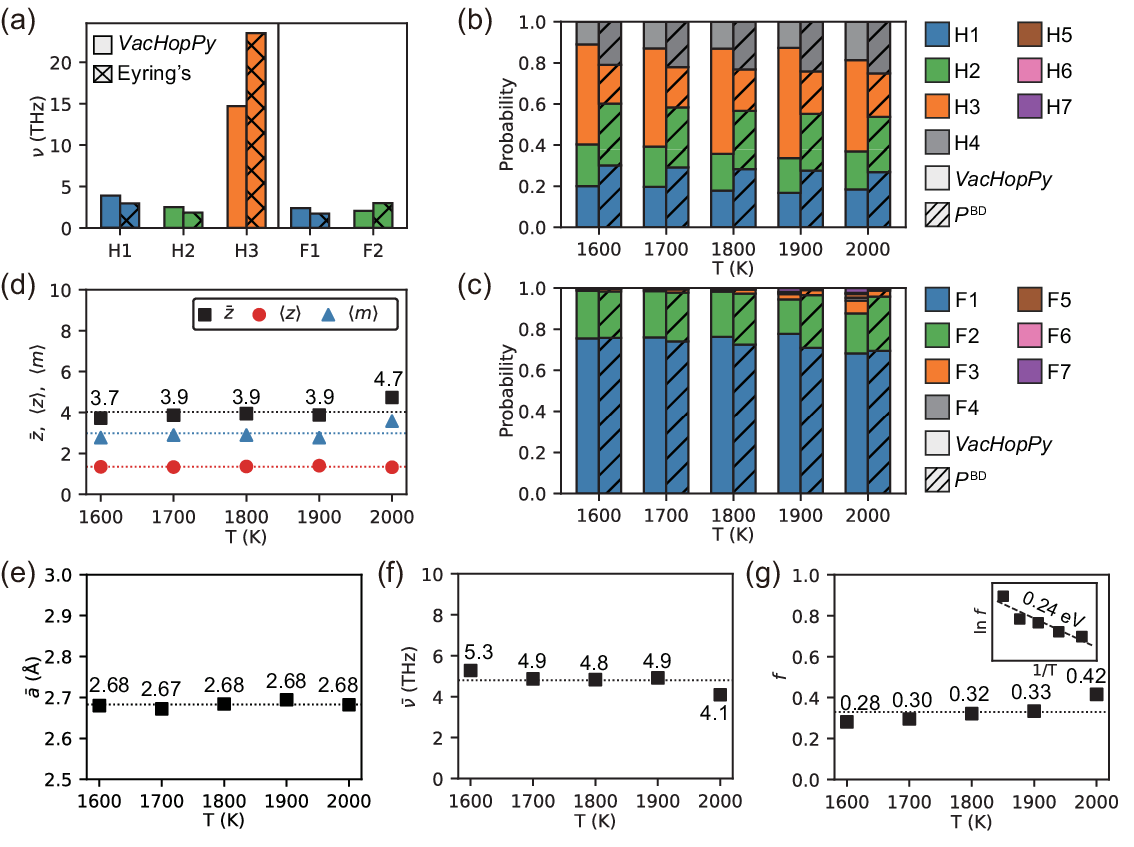}
  \caption{\textit{VacHopPy} results for \textit{mono} \ce{HfO2}. (a) Averaged values of $\nu$ over the simulated $T$ range for the five major paths. For comparison, the corresponding values calculated using Eyring’s formula are shown as x-patterned bars. (b) and (c) Pathwise hopping probabilities for the H-paths and F-paths, respectively. Solid bars represent the \textit{VacHopPy} results that account for vibrational anisotropy, while hatched bars indicate Boltzmann-weighted probabilities ($P^\mathrm{BD}$) that neglect anisotropy. (d) $T$-dependence of $\bar{z}$, $\langle z \rangle$, and $\langle m \rangle$. (e) $T$-dependence of $\bar{a}$. (f) $T$-dependence of $\bar{\nu}$.(g) $T$-dependence of $f$. The inset shows an Arrhenius plot of $f$ with the best linear fit indicated by a dashed line. In panels (d)-(g), dotted lines represent the mean values over the simulated $T$ range.}
  \label{fig8}
\end{figure}

Fig.~\ref{fig8}(a) shows the values of $\nu$ for paths with low $E_a$ (H1, H2, and H3 for the H-site; F1 and F2 for the F-site) averaged over the simulated $T$ range. For comparison, the calculations were conducted using two methods: \textit{VacHopPy} (solid bars) and Eyring’s formula (x-patterned bars; see Supplementary Note~2). Both methods show similar trends, with the H3 path exhibiting a significantly higher $\nu$ compared to the others, suggesting the thermal vibration at the H-site is strongly biased toward the H3 path. The slight discrepancies between the two methods may originate from the harmonic approximation used in phonon calculations or from the presence of imaginary phonon modes in \textit{mono} \ce{HfO2} (see Fig.~N10 of the Supplementary Material).

To examine the effect of anisotropic vibration, the probability of vacancy hopping through each path was evaluated both with and without considering vibrational anisotropy. The former was directly obtained from \textit{VacHopPy}, while the latter was computed using a Boltzmann-weighted distribution, $P_{(i,j)}^{\mathrm{BD}} \propto z_{(i,j)} P_{(i,j)}^{\mathrm{esc}}$, which excludes the factor $\nu_{(i,j)}$. This expression is widely used in other frameworks, such as KMC simulations~\cite{ref16,ref50}. Figs.~\ref{fig8}(b) and (c) show the pathwise hopping probabilities for the H-paths and F-paths, respectively. For clarity, the probabilities for H- and F-paths are presented separately, though overall, the contribution from H-paths is approximately three times greater than that from F-paths (3,049 vs.\ 987 events). For the H-paths, the probability of hopping via the H3 path from \textit{VacHopPy} far exceeds the corresponding $P^{\mathrm{BD}}$ values across the simulated $T$ range, indicating that vibrational anisotropy plays a crucial role at the H-site. In contrast, for the F-site, the probabilities from \textit{VacHopPy} closely match the $P^{\mathrm{BD}}$ values, suggesting that vibrational anisotropy has a negligible effect in this case.

The anisotropic vibration biased toward the H3 path likely originates from the lowest-energy optical phonon. Fig.~S2 of the Supplementary Material shows the phonon dispersion of \textit{mono} \ce{HfO2} along with the vibrational mode corresponding to the lowest-energy optical phonon at the $\Gamma$ point. In this mode, the displacements of the oxygen atoms are approximately aligned with the H3 path. Since lower-energy phonons have higher thermal occupations, thermal vibrations are biased toward the H3 path. This mode is therefore expected to play a significant role not only in vacancy-mediated transport but also in structural transformation; in fact, modulating the structure along its eigenvector induces a transition from the \textit{mono} to the \textit{tet} phase, as described in Fig.~S2(c). Moreover, this phonon mode can drive a collective migration of oxygen atoms along the $c$-axis, with displacements coherently following the H3 path, as shown in Fig.~S3 of the Supplementary Material, further emphasizing its relevance to atomic transport in \textit{mono} \ce{HfO2}.

The effective hopping parameters for \textit{mono} \ce{HfO2} are summarized in Table~\ref{tab:parameters_hfo2}. As shown in Figs.~\ref{fig8}(b) and (c), five major paths, H1, H2, H3, F1, and F2, dominate the overall hopping process. Considering their relative contributions, $\bar{a}$ and $\bar{E}_a$ appear reasonable, as they are comparable to or slightly lower than the corresponding values for the H3 path, the most dominant one. Notably, the calculated $\bar{E}_a$ of 0.68~eV is in good agreement with experimental measurements (0.67–0.72~eV, 0.48~eV) and MD simulations (0.66~eV)~\cite{ref51,ref52,ref53}.

\begin{table}[htbp]
  \renewcommand{\arraystretch}{1.4}
  \setlength{\tabcolsep}{4pt}
  \caption{Effective hopping parameters for \ch{V_O^{2+}} in \textit{mono} \ce{HfO2} ($T=1600-2000$ K). For $T$-dependent parameters, the mean value over the simulated $T$ range is represented, with the standard deviation given in parentheses.}
  \label{tab:parameters_hfo2}
  \centering
  \footnotesize
  \begin{tabular}{@{}>{\centering\arraybackslash}p{2cm} >{\centering\arraybackslash}p{4.8cm}@{}}
    \toprule
    \multicolumn{2}{c}{\textbf{Effective hopping parameter (\ce{HfO2})}}\\
    \midrule
    \multicolumn{2}{c}{\textit{Path parameters}} \\
    \cmidrule(lr){1-2}
    $\bar{a}$       & 2.68 ($\pm$0.01) Å \\
    $\bar{E}_a$     & 0.68 eV \\
    $\bar{z}$       & 4.0 ($\pm$0.4) \\
    $\bar{\nu}$     & 4.8 ($\pm$0.4) THz \\
    \cmidrule(lr){1-2}
    \multicolumn{2}{c}{\textit{Transport parameters}} \\
    \cmidrule(lr){1-2}
    $\bar{D}_0$     & $3.7 \times 10^{-7}$ m$^2$/s \\
    $\bar{\tau}_0$  & $5.2 \times 10^{-2}$ ps \\
    $f$             & 0.33 ($\pm$0.05) \\
    \bottomrule
  \end{tabular}
\end{table}

Fig.~\ref{fig8}(d) shows $\bar{z}$ (black squares) as a function of $T$, with its average value indicated by the black dotted line. While $\bar{z}$ remains nearly constant up to 1900~K, a noticeable increase is observed at 2000~K. To investigate the origin of this increase, $\bar{z}$ is factorized as $\bar{z} = \langle z \rangle \cdot \langle m \rangle$, where $\langle z \rangle$ represents the mean number of equivalent paths per path type, and $\langle m \rangle$ is the mean number of path types participating in the hopping process. The expressions for $\langle z \rangle$ and $\langle m \rangle$ are given by Eqs.~\eqref{eqA15} and \eqref{eqA16}, respectively. In Fig.~\ref{fig8}(d), while $\langle z \rangle$ (red circles) remains nearly constant with increasing $T$, $\langle m \rangle$ (blue triangles) exhibits an abrupt increase at 2000~K, clearly indicating that the rise in $\bar{z}$ is primarily due to the activation of additional path types at higher $T$. This observation is consistent with the path distributions shown in Figs.~\ref{fig8}(b) and (c), where more diverse hopping paths are activated at 2000~K. Meanwhile, Figs.~\ref{fig8}(e) and (f) show $\bar{a}$ and $\bar{\nu}$ as functions of $T$, with their mean values indicated by dotted lines. The $\bar{a}$ value remains nearly constant with increasing $T$, whereas $\bar{\nu}$ decreases. This decrease in $\bar{\nu}$ is consistent with Fig.~\ref{fig8}(b), where the fraction of the H3 path—the path with the largest $\nu$—decreases as $T$ rises.

Fig.~\ref{fig8}(g) shows $f$ as a function of $T$, with its mean value indicated by a dotted line. The inset represents the Arrhenius fit. Unlike the negligible $T$-dependence of $f$ observed in fcc~Al and rutile~\ce{TiO2}, a pronounced increase is observed. This behavior is likely related to the instability of the \textit{mono} lattice near the onset of a phase transition to the \textit{tet} phase, which is known to occur around 2000~K. Near this transition temperature, the coexistence of \textit{mono} and \textit{tet} phases may lead to ambiguous definition of lattice sites, thereby increasing the randomness of atomic motion and consequently raising the value of $f$. From the Arrhenius fit, the activation energy associated with $f$ ($E_a^f$) is estimated to be 0.24~eV. This factor should be taken into account when modeling the $T$-dependence of $D$ in this $T$ regime, as described by Eqs.~\eqref{eq1} and \eqref{eq6}.

Compared to fcc~Al and rutile~\ce{TiO2}, the $f$ in \textit{mono}~\ce{HfO2} is relatively lower, implying that \ch{V_O^{2+}} migration is highly correlated. A major reason for this strong correlation is the relationship between the H1 and F1 paths: the H1 path corresponds to the hopping from the H-site to the F-site, while the F1 path is the reverse process of the H1 path (see Fig.~S4(a) of the Supplementary Material). Considering the significantly lower $E_a$ of the F1 path and the higher energetic state of the F-site (by 0.56~eV), a \ch{V_O^{2+}} that hops via the H1 path tends to quickly return via the F1 path, indicating a pronounced back-and-forth correlation. This is quantitatively supported by Fig.~S4(b), which shows the return probabilities (the likelihood that a \ch{V_O^{2+}} returns to its original site after a hop) for the major paths. As expected, the H1 path exhibits a return probability of 76\%, more than twice those of the other paths.

\subsubsection{High-$T$ region ($T > 2000$~K)}
\label{sec5.3.2}

At $T > 2000$~K, the \textit{mono}~\ce{HfO2} becomes unstable and undergoes a phase transition to the \textit{tet}~\ce{HfO2}. In such unstable lattices, the center of atomic vibration may shift away from the original lattice sites, resulting in poorly defined vacancy sites. Consequently, the determination of vacancy trajectories and the calculation of effective hopping parameters may lose accuracy. The structural stability of \textit{mono}~\ce{HfO2} was evaluated using the \texttt{\footnotesize distance} command with the arguments of $t_\mathrm{interval}=0.07$ ps, $R_\mathrm{max}=20$ Å, $\Delta = 0.04$ Å, and $\sigma=0.04$ Å (see Appendix~C for details).

\begin{figure}[htbp]
  \centering
  \includegraphics[width=\textwidth]{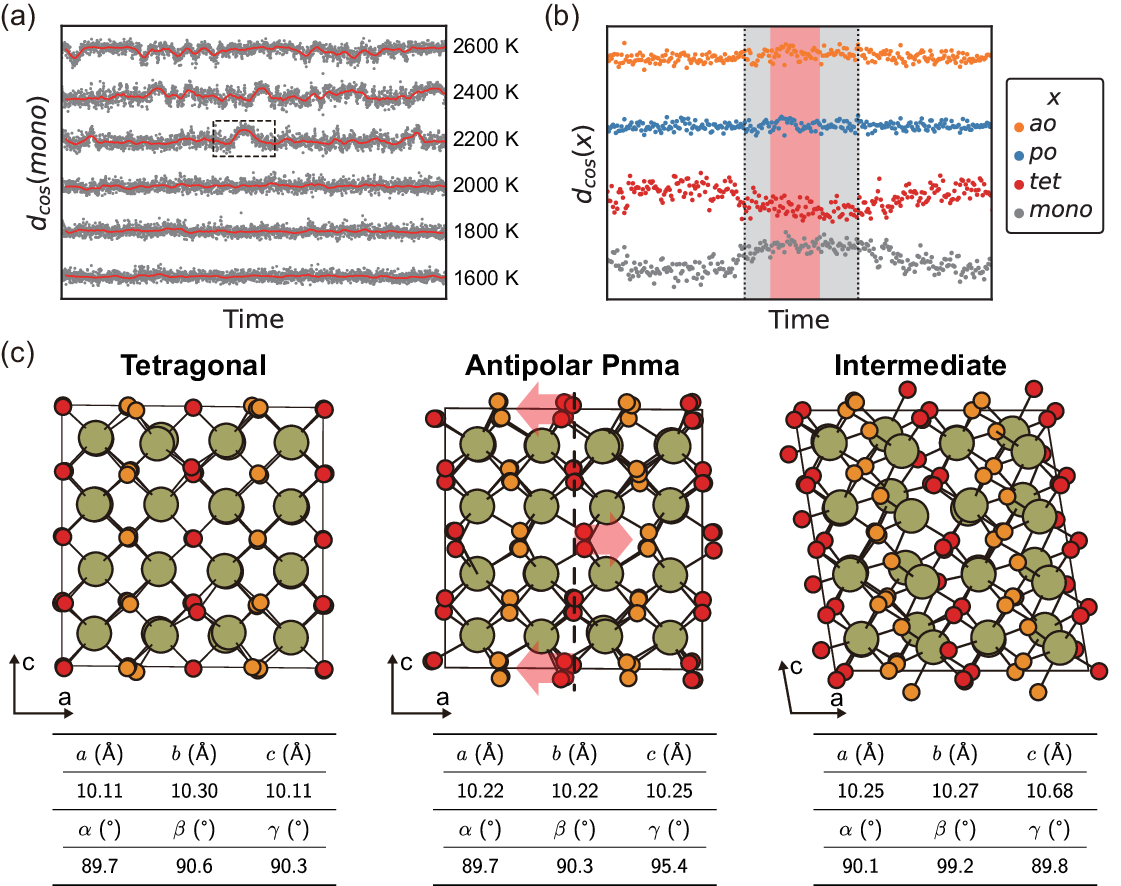}
  \caption{(a) Time evolution of $d_\mathrm{cos}(mono)$ at each simulated $T$ for \textit{mono}~\ce{HfO2}, with the simple moving averages (SMAs) represented as the red lines. For comparison, the traces at each $T$ are vertically offset; hence, the absolute $y$-values are meaningless. The key information is the relative variations in $d_\mathrm{cos}(mono)$ at each $T$. (b) Magnified view of the peak observed at 2200~K, indicated by a dashed box in (a), with additional comparisons to three reference phases (\textit{x}): antipolar orthorhombic (\textit{ao}), polar orthorhombic (\textit{po}), and tetragonal (\textit{tet}). (c) Three representative structures obtained by fully relaxing snapshots within the peak region (enclosed by two dashed lines in (b)): a \textit{tet}-like structure (left), an antipolar orthorhombic Pnma-like structure (middle), and an intermediate (\textit{im}) structure (right). All structures contain a \ch{V_O^{2+}}, resulting in slight deviations in lattice parameters from their ideal phases. In the middle panel, thick red arrows indicate the displacement directions of oxygen atoms, highlighting the antipolar nature of the Pnma-like structure.}
  \label{fig9}
\end{figure}

Fig.~\ref{fig9}(a) shows the time evolution of $d_\mathrm{cos}(mono)$ during AIMD simulations at three low-$T$ (1600~K, 1800~K, and 2000~K) and three high-$T$ (2200~K, 2400~K, and 2600~K). To clarify the overall trend, the simple moving average (SMA) of each trace is also plotted as a red line. The traces at each $T$ are vertically offset, and thus, the absolute $y$-values are meaningless. Instead, the focus is on the relative changes in $d_\mathrm{cos}(mono)$ over time. In the low-$T$ regime, $d_\mathrm{cos}(mono)$ remains nearly constant, indicating structural stability of the \textit{mono} phase. In contrast, the high-$T$ regime exhibits substantial fluctuations, suggesting increased lattice instability and the onset of the phase transition. It is worth noting that the lattice parameters were constrained to those of the \textit{mono} phase during the AIMD simulations, as they were conducted under the NVT ensemble. As a result, any structural deformation is not sustained, and the system tends to revert to the \textit{mono} structure, leaving a peak in the $d_\mathrm{cos}(mono)$ trace.

Fig.~\ref{fig9}(b) shows a magnified view of the peak region indicated by the dashed box in Fig.~\ref{fig9}(a). For comparison, three representative phases are additionally considered as reference phases: antipolar orthorhombic (\textit{ao}, Pbca), polar orthorhombic (\textit{po}, Pca2$_1$), and \textit{tet}. Their atomic structures are provided in Fig.~S5 of the Supplementary Material. Notably, as $d_\mathrm{cos}$ (\textit{mono}) increases, $d_\mathrm{cos}(tet)$ decreases, while $d_\mathrm{cos}(ao)$ and $d_\mathrm{cos}(po)$ remain nearly constant. This trend clearly suggests that the phase transition is directed toward the \textit{tet} phase.

To further confirm this observation, a total of 81 structural snapshots within the peak region, enclosed by the two vertical dashed lines in Fig.~\ref{fig9}(b), were fully relaxed. Among them, only four snapshots returned to the initial \textit{mono} phase, while 42 and 7 snapshots relaxed into the \textit{tet} phase and the antipolar orthorhombic Pnma phase, respectively. The Pnma phase can be obtained by deforming the \textit{tet} phase~\cite{ref54}, and its structure is described in Fig.~S5(e) and Fig.~S6 of the Supplementary Material. The remaining 28 snapshots relaxed into an unfamiliar structure, hereafter referred to as the intermediate (\textit{im}) structure. Its structural, energetic, and dynamical properties are described in Fig.~S7 of the Supplementary Material. The \textit{im} structure exhibits the lowest energy among all phases considered, except for the \textit{mono} phase. However, it also shows a weak imaginary phonon mode at a non-high-symmetry point, indicating dynamic instability. The three representative relaxed structures are shown in Fig.~\ref{fig9}(c). As all these snapshots contain a \ch{V_O^{2+}}, their lattice parameters deviate slightly from those of the ideal phase (axis transformations are needed for direct comparison).

Interestingly, the snapshots that relaxed into the \textit{tet} phase are predominantly located in the central peak region (red background in Fig.~\ref{fig9}(b)), whereas those that relaxed into the Pnma or \textit{im} structure are located outside the peak region in the grey shade. This distribution suggests that the \textit{im} structure represents an intermediate stage or a by-product in the transformation between the \textit{mono} and the \textit{tet} phases. Similar to the \textit{mono} phase, the \textit{im} structure can be transformed into the \textit{tet} phase by the modulation along the eigenvector of its lowest optical phonon mode, where the detailed procedure is summarized in Fig.~S7(d).

In summary, the \texttt{\footnotesize distance} command serves as a versatile diagnostic tool for MD simulations. It enables (1) validation of lattice site stability, allowing reliable tracking of vacancies; (2) identification of phase transitions through structural deviations; and (3) discovery of novel or metastable configurations that may emerge during the simulation.

\subsection{Applicability of \textit{VacHopPy}}
So far, the effective hopping parameters for three example systems have been computed using \textit{VacHopPy}. In our framework, the transport parameters $D_\mathrm{rand}$ and $\tau$ are approximated by simple Arrhenius expressions, $\bar{D}_\mathrm{rand}$ and $\bar{\tau}$, as described in Eqs.~\eqref{eq1} and \eqref{eq2}. However, for systems with multiple hopping paths, the accurate expressions for $D_\mathrm{rand}$ and $\tau$ consist of multiple exponential terms, as described in Section~2.2. This complexity raises concerns about the validity of the Arrhenius-type approximation. Fig.~S8 of the Supplementary Material shows comparisons between $D_\mathrm{rand}$ and $\bar{D}_\mathrm{rand}$, and between $\tau$ and $\bar{\tau}$, for three example systems, along with their coefficient of determination ($R^2$). In all cases, including \textit{mono}~\ce{HfO2} having 14 distinct hopping paths, good agreement is observed. This demonstrates the broad applicability of our framework across various crystalline systems and confirms that the effective hopping parameters remain valid over a wide $T$ range spanning several hundred kelvins. Moreover, \textit{VacHopPy} provides raw diffusivity data at each $T$, allowing users to fit the results to user-defined models when a single Arrhenius equation is insufficient.

\textit{VacHopPy} is also compatible with systems containing multiple vacancies. The \texttt{\footnotesize trajectory} command visualizes the individual diffusion trajectories of each vacancy (see Fig.~S9 of the Supplementary Material), and the \texttt{\footnotesize analyze} command provides various effective hopping parameters, except for $\bar{z}$ and $\bar{\nu}$, which are currently supported only for monovacancy systems. However, the reliability of \textit{VacHopPy} may decrease as the vacancy concentration increases. This potential limitation arises because \textit{VacHopPy} assumes fixed lattice sites. In regions with a high local vacancy concentration, the equilibrium position of an atom (\textit{i.e.}, its vibration center) can deviate from its original site, as shown in Fig.~S10 of the Supplementary Material, thereby increasing the likelihood of misidentifying hopping events.

\textit{VacHopPy} also supports MSD-based $D$ calculations using the \texttt{\footnotesize msd} command, which follows the approach commonly adopted in existing codes~\cite{ref15,mdanalysis,kinisi}. As shown in Fig.~S11 of the Supplementary Material, this method yields results comparable to those obtained from the \textit{VacHopPy} framework. While this approach is applicable to general diffusion mechanisms, it cannot provide hopping parameters other than $D$.

To ensure reliable \textit{VacHopPy} implementation, sufficient sampling of hopping events is essential. Empirically, statistical convergence is generally attained with approximately 300-500 events at each temperature, although the precise number may vary depending on the system. Although this study employed AIMD for sampling hopping events, a more practical alternative is machine-learning interatomic potential (MLIP)-based MD simulation, as it enables larger-scale simulations at significantly reduced computational cost~\cite{ref55,ref56}.

\section{Conclusion}
We present \textit{VacHopPy}, an open-source Python package that provides a robust and automated framework for analyzing vacancy hopping mechanisms based on MD simulations. Designed for multiscale modeling, \textit{VacHopPy} extracts a complete set of effective hopping parameters in a simplified and consistent form suitable for continuum-scale models.

These effective parameters are statistically derived by integrating energetic, kinetic, and geometric contributions from all hopping paths, thereby accurately capturing material-specific hopping behavior. Particularly, vibrational anisotropy, often overlooked in other frameworks, is inherently considered and shown to play a crucial role in determining diffusion characteristics. Furthermore, this package enables \textit{ab initio}-based calculation of the correlation factor, $f$, allowing the effects of lattice instability and other complex behaviors to be reflected. The applicability and accuracy of \textit{VacHopPy} are demonstrated using three representative systems: \textit{fcc}~Al (with a single site and a single hopping path), rutile~\ce{TiO2} (with a single site and multiple hopping paths), and \textit{mono}~\ce{HfO2} (with multiple sites and multiple hopping paths). In all cases, the extracted effective parameters show good agreement with previously reported experimental and theoretical results.

Beyond parameter extraction, \textit{VacHopPy} offers tools for tracking vacancy trajectories and detecting phase transitions during MD simulations. For instance, in \textit{mono}~\ce{HfO2}, it captures the collective oxygen migration along the $c$-axis and the onset of a tetragonal phase transition at 2200~K, offering deeper insights into atomic-scale transport mechanisms.

Overall, \textit{VacHopPy} establishes a general and transferable framework that bridges atomistic simulations and continuum-scale modeling, enabling more accurate and reliable multiscale simulations across a wide range of materials.

\section*{Authorship contributions}
Taeyoung Jeong: Formal analysis, Investigation, Methodology, Writing - original draft; Kun Hee Ye: Formal analysis, Writing - review \& editing; Seungjae Yoon: Formal analysis, Writing - review \& editing; Dohyun Kim: Formal analysis, Writing - review \& editing; Yunjae Kim: Formal analysis, Writing - review \& editing; Cheol Seong Hwang: Writing - review \& editing. Jung-Hae Choi: Conceptualization, Funding acquisition, Project administration, Supervision, Writing - review \& editing

\section*{Competing interest}
The authors declare that they have no known competing financial interests or personal relationships that could have appeared to influence the work reported in this paper.

\section*{Acknowledgments}
J.-H.C. was supported by the National Research Council of Science \& Technology (NST) grant by MSIT [No. GTL24041-000], by the National Research Foundation of Korea (NRF) grant by MSIT [Next Generation Intelligence Semiconductor Foundation 2022M3F3A2A01076569], and by the Institutional Research Program of the Korea Institute of Science and Technology (KIST) [2E33873]. 

\section*{Data availability}
The data that supports the findings of this study are available in the article and its Supplementary Material.

\section*{Declaration of generative AI and AI-assisted technologies in the writing process}
During the preparation of this work the authors used ChatGPT in order to improve language and readability. After using this service, the authors reviewed and edited the content as needed and take full responsibility for the content of the publication.

\appendix
\section*{Appendix A. Equations for effective hopping parameters}

Thermodynamically, the hopping rate $\Gamma_{(i,j)}$ and the diffusivity $D_{(i,j)}$ corresponding to the $(i,j)$ path are given by:
\begin{equation}
\Gamma_{(i,j)} = z_{(i,j)} \nu_{(i,j)} P_{(i,j)}^{\mathrm{esc}}, \quad \text{where} \quad P_{(i,j)}^{\mathrm{esc}} = \exp\left(-\frac{E_{a,(i,j)}}{k_B T}\right)
\tag{A.1}
\label{eqA1}
\end{equation}
\begin{equation}
D_{(i,j)} = \frac{1}{6} a_{(i,j)}^{2} \Gamma_{(i,j)}
\tag{A.2}
\label{eqA2}
\end{equation}

Assuming an ideal random walk, the random walk diffusivity, $D_\mathrm{rand}$, is expressed as a linear combination of $D_{(i,j)}$, since each hopping event is treated independently:
\begin{equation}
D_\mathrm{rand} = \sum_{i=1}^n P_i^\mathrm{site} \sum_{j=1}^{m_i} D_{(i,j)}
\tag{A.3}
\label{eqA3}
\end{equation}
\begin{equation}
= \frac{1}{6 t} \sum_{(i,j)} a_{(i,j)}^{2} c_{(i,j)}
\tag{A.4}
\label{eqA4}
\end{equation}
Here, $\sum_{(i,j)}$ represents the double summation $\sum_{i=1}^n \sum_{j=1}^{m_i}$, and $P_i^\mathrm{site}$ is the site occupation probability, thermodynamically proportional to $\exp(-E_{f,i}/(k_B T))$, where $E_{f,i}$ is the vacancy formation energy at the $i$th site. The second equality in Eq.~\eqref{eqA4} is derived by employing Eq.~\eqref{eqA2} and the relations $P_i^\mathrm{site} = t_i/t$ and $\Gamma_{(i,j)} = c_{(i,j)}/t_i$.

Similarly, the overall hopping rate ($\Gamma$) is expressed as:
\begin{equation}
\Gamma = \sum_{i=1}^n P_i^\mathrm{site} \sum_{j=1}^{m_i} \Gamma_{(i,j)}
\tag{A.5}
\label{eqA5}
\end{equation}
\begin{equation}
= \sum_{(i,j)} \nu_{(i,j)} P_{(i,j)}, \quad P_{(i,j)} = z_{(i,j)} P_i^\mathrm{site} P_{(i,j)}^{\mathrm{esc}}
\tag{A.6}
\label{eqA6}
\end{equation}

The second equality is derived using Eqs.~\eqref{eqA1}, \eqref{eqA2}, and the relation $P_i^\mathrm{site} = t_i/t$. In Eq.~\eqref{eqA6}, $P_{(i,j)}$ is proportional to the probability that a hop occurs via the $(i,j)$ path, where $z_{(i,j)}$ represents the path degeneracy. Note that these equations are not restricted to ideal random walks but are generally valid for arbitrary hopping processes.

Since $\bar{\nu}$ is defined as the representative value for $\nu_{(i,j)}$, $\nu_{(i,j)}$ in Eq.~\eqref{eqA6} can be replaced with $\bar{\nu}$, resulting in:
\begin{equation}
\Gamma = \bar{\nu} \sum_{(i,j)} P_{(i,j)}
\tag{A.7}
\label{eqA7}
\end{equation}
where $\bar{\nu}$ is factored out of the summation since it is independent of the specific hopping path. Comparing Eqs.~\eqref{eqA6} and \eqref{eqA7} yields:
\begin{equation}
\bar{\nu} = \frac{\sum_{(i,j)} \nu_{(i,j)} P_{(i,j)}}{\sum_{(i,j)} P_{(i,j)}}
\tag{A.8}
\label{eqA8}
\end{equation}
\begin{equation}
= \frac{c}{t \sum_{(i,j)} P_{(i,j)}}
\tag{A.9}
\label{eqA9}
\end{equation}

The second equality is derived by combining Eq.~\eqref{eqA6} with the relation $\Gamma = c/t$. Physically, Eq.~\eqref{eqA8} implies that $\bar{\nu}$ represents the ensemble average of $\nu_{(i,j)}$ weighted by $P_{(i,j)}$. Although Eq.~\eqref{eqA9} is implemented for the calculation of $\bar{\nu}$ in \textit{VacHopPy}, Eq.~\eqref{eqA8} is presented in the main text to clarify its physical interpretation.

While $\Gamma$ in the original lattice is represented as a sum of multiple exponential terms (see Eq.~\eqref{eqA6}), it reduces to a single exponential term in the effective lattice, where only a single effective hopping path exists. Accordingly, the effective hopping rate, $\bar{\Gamma}$, is expressed as:
\begin{equation}
\bar{\Gamma} = \bar{z} \bar{\nu} \bar{P}^{\mathrm{esc}}, \quad \bar{P}^{\mathrm{esc}} = \exp\left(-\frac{\bar{E}_a}{k_B T}\right)
\tag{A.10}
\label{eqA10}
\end{equation}
Since the effective lattice reproduces the overall hopping properties of the original lattice, the relation $\bar{\Gamma} \approx \Gamma$ must hold. Substituting Eqs.~\eqref{eqA7} and \eqref{eqA10} results in:
\begin{equation}
\bar{z} = \frac{\sum_{(i,j)} P_{(i,j)}}{\bar{P}^{\mathrm{esc}}}
\tag{A.11}
\label{eqA11}
\end{equation}

To facilitate further analysis, $\bar{z}$ is factorized into $\langle z \rangle$ and $\langle m \rangle$ as $\bar{z} = \langle z \rangle \cdot \langle m \rangle$. Here, $\langle z \rangle$ represents the mean number of equivalent paths per path type, and $\langle m \rangle$ corresponds to the mean number of path types participating in the hopping process.

To derive $\langle z \rangle$, Eq.~\eqref{eqA5} is combined with Eq.~\eqref{eqA1}, yielding:
\begin{equation}
\Gamma = \sum_{i=1}^n P_i^\mathrm{site} \sum_{j=1}^{m_i} z_{(i,j)} \nu_{(i,j)} P_{(i,j)}^{\mathrm{esc}}
\tag{A.12}
\label{eqA12}
\end{equation}
By definition, $z_{(i,j)}$ is replaced with $\langle z \rangle$, leading to:
\begin{equation}
\Gamma = \langle z \rangle \sum_{i=1}^n P_i^\mathrm{site} \sum_{j=1}^{m_i} \nu_{(i,j)} P_{(i,j)}^{\mathrm{esc}}
\tag{A.13}
\label{eqA13}
\end{equation}
The term $\langle z \rangle$ is taken out of the summation because it is irrelevant to path selection.

Now, by combining Eq.~\eqref{eqA1} with the relation $\Gamma_{(i,j)} = c_{(i,j)} / t_i$, the following expression is obtained:
\begin{equation}
\nu_{(i,j)} P_{(i,j)}^{\mathrm{esc}} = \frac{c_{(i,j)}}{z_{(i,j)} t_i}
\tag{A.14}
\label{eqA14}
\end{equation}
Substituting Eq.~\eqref{eqA14} into Eq.~\eqref{eqA13} and applying the relations $\Gamma = c / t$ and $P_i^\mathrm{site} = t_i / t$, we derive:
\begin{equation}
\langle z \rangle = \frac{c}{\sum_{(i,j)} \frac{c_{(i,j)}}{z_{(i,j)}}}
\tag{A.15}
\label{eqA15}
\end{equation}
Hence, $\langle z \rangle$ becomes the harmonic mean of $z_{(i,j)}$ weighted by $c_{(i,j)}$, and $\langle m \rangle$ is then obtained as:
\begin{equation}
\langle m \rangle = \frac{\bar{z}}{\langle z \rangle}
\tag{A.16}
\label{eqA16}
\end{equation}

Meanwhile, by rearranging Eq.~\eqref{eqA14}, the attempt frequency for the $(i,j)$ path, $\nu_{(i,j)}$, is expressed as:
\begin{equation}
\nu_{(i,j)} = \frac{c_{(i,j)}}{z_{(i,j)} t_i P_{(i,j)}^{\mathrm{esc}}}
\tag{A.17}
\label{eqA17}
\end{equation}
Consequently, $\nu_{(i,j)}$ is statistically determined using this equation in \textit{VacHopPy}, contrasting with conventional methods based on TST that rely on phonon properties. This statistical representation is particularly advantageous for systems with unstable phonon branches, where TST-based methods often become unreliable.

\section*{Appendix B. Equations for correlation factor}

Using the Einstein relation, $\langle \mathbf{R}^2 \rangle = 6 D t$, Eq.~\eqref{eq6} can be rewritten as:
\begin{equation}
f = \frac{\langle \mathbf{R}^2 \rangle}{\langle \mathbf{R}_{\mathrm{rand}}^2 \rangle}
\tag{B.1}
\label{eqB1}
\end{equation}
where $\langle \mathbf{R}^2 \rangle$ and $\langle \mathbf{R}_{\mathrm{rand}}^2 \rangle$ represent the mean squared displacements (MSD) of the actual diffusion process and an ideal random walk process, respectively.

According to random walk theory, $\langle \mathbf{R}_{\mathrm{rand}}^2 \rangle$ is given by:
\begin{equation}
\langle \mathbf{R}_{\mathrm{rand}}^2 \rangle = \sum_{k=1}^c \langle \mathbf{r}_k^2 \rangle
\tag{B.2}
\label{eqB2}
\end{equation}
where $\mathbf{r}_k$ is the displacement of the $k$th hop, and $c$ is the total number of hopping events. Let $c_{(i,j)}$ be the number of hops via the $(i,j)$ path. Then, $\langle \mathbf{R}_{\mathrm{rand}}^2 \rangle$ can be expressed as:
\begin{equation}
\langle \mathbf{R}_{\mathrm{rand}}^2 \rangle = \sum_{(i,j)} a_{(i,j)}^{2} c_{(i,j)}
\tag{B.3}
\label{eqB3}
\end{equation}
\begin{equation}
= c \sum_{(i,j)} a_{(i,j)}^{2} q_{(i,j)}, \quad q_{(i,j)} = \frac{c_{(i,j)}}{c}
\tag{B.4}
\label{eqB4}
\end{equation}
Here, $q_{(i,j)}$ represents the probability that a hop occurs via the $(i,j)$ path. Note that Eqs.~\eqref{eqA4} and \eqref{eqB3} satisfy the Einstein relation.

Combining Eqs.~\eqref{eqB1} and \eqref{eqB4}, we obtain:
\begin{equation}
f = \frac{\langle \mathbf{R}^2 \rangle}{c \sum_{(i,j)} a_{(i,j)}^{2} q_{(i,j)}}
\tag{B.5}
\label{eqB5}
\end{equation}
This expression corresponds to Eq.~\eqref{eq7} of the main text.

To improve computational convergence, \textit{VacHopPy} employs the \textit{encounter} approach. An \textit{encounter} is defined as a sequence of exchanges between a tracer atom and a vacancy, which terminates when the tracer atom interacts with another vacancy. Detailed explanations for this approach can be found elsewhere~\cite{ref58,ref59}. To adopt the \textit{encounter} framework, Eq.~\eqref{eqB5} is slightly modified as follows:
\begin{equation}
f = \frac{\langle \mathbf{R}^2 \rangle_\mathrm{enc}}{c_\mathrm{enc} \sum_{(i,j)} a_{(i,j)}^{2} q_{(i,j)}}
\tag{B.6}
\label{eqB6}
\end{equation}
Here, $\langle \mathbf{R}^2 \rangle_\mathrm{enc}$ represents the MSD per \textit{encounter}, and $c_\mathrm{enc}$ is the average number of tracer-vacancy exchanges per \textit{encounter}. In \textit{VacHopPy}, vacancies in periodic image cells are treated individually: when a tracer atom exchanges with a vacancy in the original cell and subsequently interacts with another vacancy in a neighboring image cell, the original \textit{encounter} ends, and a new \textit{encounter} begins. Accordingly, even when only a single vacancy exists in the input cell, $c_\mathrm{enc}$ is usually in the range of 1 to 2. This approach allows a number of \textit{encounter}s to be sampled from MD, thereby enhancing the convergence of $f$.

\section*{Appendix C. Fingerprint analysis}

To obtain a fingerprint vector ($\psi$) for atomic configurations, also known as a descriptor, we employ an expression proposed by Oganov \textit{et al}.~\cite{ref60}. The fingerprint vector corresponding to a pair of atom types $A$ and $B$ ($\psi_{AB}$) is given by:
\begin{equation}
\psi_{AB}(\rho) = \sum_{A_i \in \text{cell}} \sum_{B_j} \left( \frac{\delta(\rho - \rho_{i,j}) \cdot V}{4 \pi \rho_{i,j}^2 N_A N_B \Delta} - 1 \right)
\tag{C.1}
\label{eqC1}
\end{equation}
Here, $\rho_{i,j}$ is the interatomic distance between atom $A_i$ and atom $B_j$. The inner summation runs over all $B_j$ atoms within a cutoff distance ($R_{\max}$) from the $A_i$ atom. $N_A$ and $N_B$ are the numbers of atoms of type $A$ and $B$, respectively. $V$ is the unit cell volume, and $\Delta$ is the bin size. The function $\delta$ is a Gaussian-smeared delta function with standard deviation $\sigma$, introduced to improve numerical convergence.

In \textit{VacHopPy}, the fingerprint of a structure is obtained by concatenating all $\psi_{AB}$ vectors corresponding to every pairwise combination of atom types $A$ and $B$. This fingerprint is invariant under translation, rotation, and mirror operations, making it a robust descriptor of atomic configurations.

Using Eq.~\eqref{eqC1}, $\psi_{AB}$ can be computed for each snapshot in an AIMD simulation. Changes in atomic configurations are quantified using the cosine distance ($d_{\mathrm{cos}}$), defined as:
\begin{equation}
d_{\mathrm{cos}}(x) = \frac{1}{2} \left( 1 - \frac{\psi_s \cdot \psi_x}{|\psi_s| \, |\psi_x|} \right)
\tag{C.2}
\label{eqC2}
\end{equation}
where $\psi_s$ is the fingerprint of a simulation snapshot, and $\psi_x$ is that of a reference phase $x$. The $d_{\mathrm{cos}}(x)$ ranges from 0 to 1, where a smaller $d_{\mathrm{cos}}(x)$ value indicates greater structure similarity between the snapshot and the reference structure.

\bibliographystyle{elsarticle-num}
\bibliography{references}

\end{document}